\documentclass[iop]{emulateapj}
\usepackage{amsmath}
\usepackage{graphicx}
\graphicspath{{Img/}}

\usepackage{enumerate}
\usepackage{tabularx}
\shorttitle{Blackbodies in GRB~090618} \shortauthors{Basak et al.}

\begin{document}

\title{DISCOVERY OF SMOOTHLY EVOLVING BLACKBODIES IN THE EARLY AFTERGLOW OF GRB 090618 :
\mbox{EVIDENCE FOR A SPINE-SHEATH JET?}}

\author{Rupal Basak$^{\rm 1,2}$ and A.R. Rao$^{\rm 2}$}

\affil{$^{\rm 1}$ Nicolaus Copernicus Astronomical Center, ul. Bartycka 18, 00-716 Warsaw, Poland \newline
$^{\rm 2}$Tata Institute of Fundamental Research, Mumbai - 400005, India. \newline
$rupal@camk.edu.pl$, $arrao@tifr.res.in$}

\begin{abstract}
GRB~090618 is a bright GRB with multiple pulses. It shows evidence of thermal 
emission in the initial pulses as well as in the early afterglow phase. As high resolution 
spectral data of \emph{Swift}/XRT is available for the early afterglow, we investigate 
the shape and evolution of the thermal component in this
phase using data from the \emph{Swift}/BAT, the \emph{Swift}/XRT, and the \emph{Fermi}/GBM detectors. 
An independent fit to the BAT and XRT data reveals two correlated blackbodies with monotonically 
decreasing temperatures. Hence we investigated the combined data with a model consisting 
of two blackbodies and a power-law (2BBPL), a model suggested for several bright 
GRBs. We elicit the following interesting features of the 2BBPL model: a) the same model 
is applicable from the peak of the last pulse in the prompt emission to the afterglow emission, 
b) the ratio of temperatures and the fluxes of the two black bodies remains constant 
throughout the observations, c) the black body temperatures and fluxes show a monotonic 
decrease with time, with the BB fluxes dropping about a factor of two faster than that
of the power-law emission, d) attributing the blackbody emission to photospheric emissions, 
we find that the photospheric radii increase very slowly with time, and the lower 
temperature blackbody shows a larger emitting radius than that of the
higher temperature black body. We find some evidence that the underlying shape of the
non-thermal emission is a cut-off power-law rather than a power-law. We sketch
a spine-sheath jet model to explain our observations.

\end{abstract}


\section{Introduction}
The radiation mechanism of the prompt emission of gamma-ray bursts (GRBs) remains an open 
question. Though the spectrum is phenomenologically fitted with the empirical Band function
(\citealt{Bandetal_1993}) having a low energy photon index ($\alpha$), a high energy photon index ($\beta$)
a peak energy ($E_{\rm peak}$) and a normalization, a physical model to describe the variety of prompt emission 
data is yet to be established. It is widely believed that the prompt emission 
spectrum is fully non-thermal, and represents an optically thin synchrotron emission 
(e.g., \citealt{Meszarosetal_1994_prompt, Piran_1999_review, Zhang_Meszaros_2002_prompt}). 
The electrons radiating the synchrotron photons are expected to cool fast, and the photon index 
of the spectrum is restricted to be less than -1.5 in this fast cooling regime (\citealt{Cohenetal_1997}; however see 
\citealt{Uhm_Zhang_2014} who suggest that an index up to -1 is possible in a moderately fast cooling regime). 
The value of $\alpha$, however, is not always compatible with 
this non-thermal model, and sometimes is found to be even greater than the maximum value allowed 
in a slow cooling regime, -2/3 (\citealt{Crideretal_1997, Crideretal_1999, Preeceetal_1998}). In addition, evidence for a thermal 
component has been seen in a few bright GRBs (e.g., \citealt{Ghirlandaetal_2003, Ryde_2004, 
Ryde_2005, Shirasakietal_2008, Ryde_Pe'er_2009, Guiriecetal_2011, Guiriecetal_2013, Axelssonetal_2012, 
Basak_Rao_2013_parametrized, Raoetal_2014}). The spectrum contains either only a thermal component,
or a combination of a thermal and a non-thermal component. The non-thermal part is
generally described by a power-law (PL) function with a photon index, $\Gamma$, 
and it represents all possible dissipative processes including an optically thin synchrotron emission.
On the other hand, the thermal component possibly signifies emission from the jet photosphere.

Such observations of thermal components during the prompt emission phase, however, 
are limited to a handful of GRBs. Moreover, the correct shape of the thermal component 
is debatable. For example, the time-resolved spectra of GRB 090902B are consistent with a variety of models ---
a single blackbody with a power-law (BBPL; \citealt{Zhangetal_2011}), a multicolour blackbody along with a 
power-law (mBBPL; \citealt{Rydeetal_2010_090902B}), and two blackbodies with a power-law 
(2BBPL; \citealt{Basak_Rao_2013_linger, Raoetal_2014}). Such a diversity in the spectral models is perhaps due 
to the limited energy resolution and/or bandwidth of the instruments used for GRB spectral analysis.
For example, the \emph{Swift}/Burst Alert Telescope (BAT) has an energy resolution of $\sim7$\,keV
(average FWHM in 15-150\,keV). Further, due to the wide field of view, 
GRB detectors are background dominated (see e.g., \citealt{Shawetal_2003}).
These constraints are particularly severe at the lower energies
(below $\sim15$\,keV) and hence reducing the effective bandwidth and making the measurement of 
multiple spectral features quite difficult.

Modelling the prompt emission spectrum is also constrained by the 
 fast spectral variability: both the flux and the spectrum of GRBs are known 
to be highly variable. Hence the average spectral shape may not
capture the inherent spectral components. On the other hand, dividing the   data into finer
time bins makes the spectral sensitivity quite poor, making it difficult to 
identify the correct spectral model. For example, for GRB~090902B, one of the brightest
GRBs, \cite{Zhangetal_2011} found that the spectrum becomes progressively narrower
when one goes down to a time bin of 0.5 s, but the spectral parameters could not be constrained
(see also \citealt{Raoetal_2014}).
One way to handle this is to smooth out the short time variability and 
capture the broad pulses, and then assume certain smooth spectral evolution within these pulses.
There is evidence that the spectral variations follow the pulse profile,
either as a hard-to-soft or intensity tracking (\citealt{LK_1996, Luetal_2012, Basak_Rao_2014_MNRAS}). 
This information can be used to examine the variations of the spectral parameter.
\cite{Basak_Rao_2013_parametrized}, for example, have adopted a new scheme for 
time-resolved spectral study, called the ``parametrized joint fit''. The procedure involves the parametrization 
of the evolution of peak energy, and tying of the spectral index in all time bins
(discussed in the context of GRB 090618 in Section 2). Such a method greatly reduces 
the number of free parameters of any spectral model, and hence, the time-resolved spectroscopy becomes more 
tractable. It is shown that the prompt emission data is consistent with two evolving blackbodies with 
a power-law (2BBPL), and the spectral evolution is quite smooth within a broad pulse of a GRB. 
However, even with this technique, only marginal conclusions could be drawn in a limited sample, namely 
GRBs with high observed flux. 

In recent years, detections of an evolving blackbody component have been reported based on the 
data from focusing X-ray detectors 
e.g., \emph{Swift}/X-Ray Telescope (XRT) (\citealt{Pageetal_2011}), \emph{NuSTAR} (\citealt{nustar_2014}), during the early
afterglow phase of a few GRBs. These detections are very convincing thanks to the impressive energy 
resolution of the X-ray detectors (e.g., 140 eV at 5.9 keV for the XRT at early time in the \emph{Swift} mission), and background reduction 
due to the focusing of X-rays. Such good quality observations are rarely done during
the prompt phase and it is not clear whether the  thermal component is an underlying
emission mechanism responsible for the bulk of the GRB emission or it is an
additional component accompanying  a predominantly non-thermal emission.
Hence it is interesting to compare the evolution of these thermal  components during the prompt and the early
afterglow phase of a GRB. With its high slewing rate, the \emph{Swift} satellite 
is capable of providing good quality data in the higher and lower energies using the BAT 
and the XRT, respectively.
Such overlapping observations, however,  are rare as the XRT generally misses the glimpse of the prompt emission, and during the 
late phase, the signal to noise of the data collected with the BAT is too low and only upper limits can be used to constrain the 
spectral shape. In addition, the XRT has a limited bandwidth (0.3-10 keV) and the blackbody needs to 
be prominent  within this bandwidth. Hence, detection of a thermal component simultaneously
with the hard non-thermal component at late prompt/ early afterglow time is a very difficult observation to make.

GRB~090618 is one rare case where a significant overlap is seen in the BAT and the XRT observations. 
This is a long GRB with $T_{90}\sim113$\,s (where $T_{90}$ is the time interval during which 
the burst emits from 5\% to 95\% of its total detected count in 50-300\,keV; \citealt{Kouveliotouetal_1993}). 
The XRT started taking the afterglow data in the Windowed Timing mode (WT) from 125\,s after the BAT trigger. 
The flux in both the detectors decreases smoothly during the overlapping observation. During the prompt emission 
phase the GRB exhibits several broad pulse emissions. \cite{Basak_Rao_2013_parametrized} have 
found evidence of two correlated blackbodies in the first two pulses. \cite{Pageetal_2011}, 
on the other hand, have found a single blackbody component in the early afterglow data of the XRT 
(primarily in the 125-275\,s interval). During the initial XRT data ($\sim125-165$\,s), the falling 
part of the last pulse is still detected in the BAT. In this context, it is interesting to investigate the thermal 
emission in the last pulse of the BAT data, and then compare its evolution with that of the XRT data. In this paper, 
we study the shape (i.e., whether it is a blackbody with a single peak, or two 
blackbodies with double peaks) and evolution 
of the thermal component during the overlap of the prompt and afterglow phase. 
In the next section (Section 2), we briefly present the data and the analysis technique.
Section 3 highlights the important timing and spectral features of this GRB. In Section 4, we 
perform a spectral analysis, and interpret the results. In Section 5, we propose a physical model to explain the 
observation. Finally, we draw conclusions and discuss the implications of our results in Section 6.

\section{Data and Analysis Technique}
We use archival data from the BAT, the XRT and the Gamma-ray Burst Monitor (GBM) on board the 
Fermi Gamma-ray Space Telescope. We use {\tt XSPEC v12.8.2} for all the spectral analyses. For the BAT and the GBM analyses, 
we follow the technique described in \cite{Basak_Rao_2012_090618}. For the joint spectral fitting with the BAT and GBM,
they have used a multiplication factor to deal with the relative calibration of the two instruments. In the present 
analysis, we essentially follow the same method. For the XRT analysis, we use the standard XRT spectrum from 
the UK Swift Science Data Centre ({\tt http://www.swift.ac.uk/burst\_analyser/}; cf. \citealt{Evansetal_2009}). 
The online software extracts the spectral data from the WT mode data with a pile-up and exposure map correction. 
We do not find any mismatch between the relative normalization of the BAT and the XRT (also see \citealt{Pengetal_2014}).
Hence, we do not use the multiplication factor for a joint BAT-XRT analysis. Keeping the relative area
of these two detectors fixed is particularly important to derive the relative importance of the two blackbodies:
the lower blackbody is predominantly seen in the XRT and the higher blackbody in the BAT. The lower energy part of the XRT
spectrum has a curvature due to a galactic and an intrinsic absorption of the source. \cite{Pageetal_2011} 
have studied the XRT spectrum of this GRB. They find the equivalent hydrogen column density for the galactic and 
source absorption as $N_{\rm H}=5.8 \times 10^{20}$ cm$^{-2}$ and $zN_{\rm H}=(1.82\pm0.08) \times 10^{21}$ cm$^{-2}$), 
respectively. We use these values for spectral fitting with the XRT.

\cite{Pageetal_2011} divide the XRT observation into four initial time bins (125-165\,s, 165-205\,s, 
205-245\,s, 245-275\,s) and one larger time bin (275-2453\,s). We use the initial time bins to study 
the evolution of the thermal component in the early afterglow phase. There are simultaneous BAT and XRT
observations during the first time bin (125-165 s), and the flux in this first time interval is considerably 
higher than that in the later ones. Hence, we further divide this interval into four equal time bins to study a finer spectral 
evolution. The time interval before the XRT observation is divided in two ways: (i) for the BAT-only analysis,
we use equal bin size, (ii) for the joint BAT-GBM analysis, we extract the time-resolved data by requiring a 
minimum of 1500 count in the NaI 4 (n4) detector, the one having the highest count rate among the GBM detectors. 
We start to integrate the count from the start of the pulse till this minimum count is achieved. Note that the 
choice of the minimum count is subjective. In order to use the BAT simultaneously with the GBM, we use $\chi^2$ 
statistics rather than C-statistics as the deconvolution technique used in the BAT instrument to extract the 
background subtracted flux gives Gaussian error rather than Poissonian error 
(see {\tt The SWIFT BAT Software Guide V6.3}). The $\chi^2$ statistics requires binning of the spectral 
channels. We choose minimum $\sim36$ counts ($6\sigma$) per spectral channel for binning, and the number 
of channels are $\sim40$, which translates into $\sim1500$ background subtracted counts per time bin. 

For the time-resolved spectroscopy, we use a new technique developed by \cite{Basak_Rao_2013_parametrized}, 
called ``parametrized joint fit''. A brief description of the method is as follows. We shall describe the method 
for a BBPL model fit to the data. However, the method is generic, and can be extended for other models.
This method implicitly assumes that GRBs consist  of pulses and the spectral parameters vary 
smoothly within these pulses at the typical times scales of the rise
and fall times of these pulses.
\cite{Ryde_Pe'er_2009} have shown that the temperature ($kT$) evolution of a blackbody has a break during the 
peak of a pulse. Based on this observation, we can divide the lightcurve of a GRB pulse into two time sectors --- 
the rising part (from the start of the pulse till the peak) and the falling part. We further divide each part,
and obtain a total of say `$n$' number of time bins. Now, a 4-parameter model like BBPL requires 
$4n$ parameters to describe the full pulse. However, with certain assumptions of the 
spectral evolution, one can reduce the number of free parameters by a large factor. The validity of the assumptions
can be checked by comparing the fit statistics with and without these assumptions. \cite{Ryde_Pe'er_2009} have shown 
that the time evolution of $kT$ in the rising and falling part of a pulse can be described by a power-law with 
two different index. Hence, 
we parameterize the evolution in each of the time sectors: $kT\sim t^{\mu}$. The ratio of the normalization of 
the blackbody and that of the power-law can be parameterized as $\sim t^{\nu}$. We also assume that the power-law 
index ($\Gamma$) remains constant in each time sector. Hence, its value can be calculated by tying 
$\Gamma$ across all the time-resolved bins of the rising and the falling part separately. With these assumptions 
we simultaneously fit all of the time-resolved spectrum in the two time sectors. Note that this new fitting scheme 
reduces the number of free parameters from $4n$ to $n+4$, e.g., for a total of 25 time bins, the number of free 
parameters reduces from 100 to 33. 

For a fitting with the 2BBPL model, in addition to tying the power-law index, we also tie the ratio of 
the temperature and normalization of the two blackbodies in all bins. This is because we find good correlation 
of these parameters among the two blackbodies. In the following we explain the method of tying the ratio of the 
temperature of the two blackbodies. Let us assume that we have three time intervals. We load all these spectra 
simultaneously in {\tt XSPEC}. Now, for the first spectrum (i.e., the spectrum for the first time bin), we give 
some initial guess values of the temperature, say $T_1(t=1)$ (higher) and $T_2(t=1)$ (lower). 
For the next spectrum, we guess an initial value for the higher temperature,
say $T_1(t=2)$. As for the lower temperature, we tie the value with ratio of 
$\left(\frac{T_2(t=1)}{T_1(t=1)}\right)\times T_1(t=2)$. The same procedure is repeated for the third bin. We 
then fit all the spectra simultaneously and obtain the best fit values of all the $T_1$ and 
that of $T_2$ for the first time bin along with the corresponding errors. The values of 
subsequent $T_2$ are trivially determined by the ratio formula above. 
Note that by tying the ratios in all bins, we assume that the parameters 
of the two blackbodies have 100\% correlation. This hypothesis can be tested by the fit statistics.

\section{Timing and Spectral Features of GRB~090618}

\begin{figure}\centering
{

\includegraphics[width=3.2in]{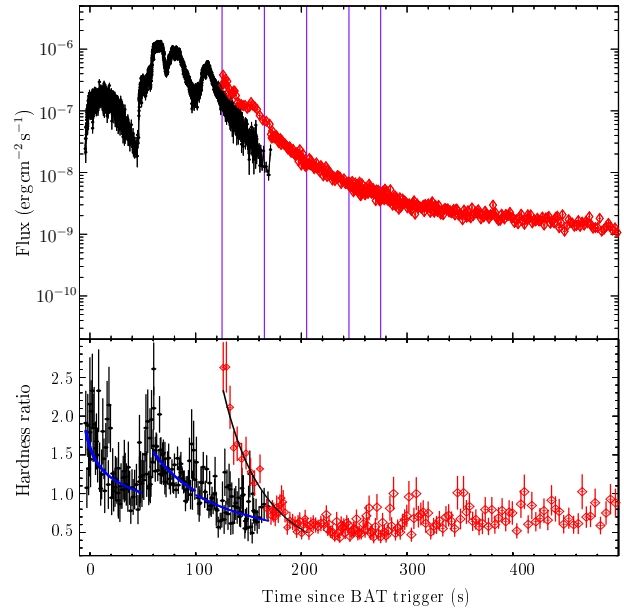} 

%\label{fig:Light Curves}
}
\caption{\emph{Upper panel:} Energy flux of GRB 090618 in the BAT (15-50\,keV) and the 
XRT (0.3-10\,keV). Filled symbols are used for the BAT and open diamonds are used for the XRT 
observations. The time intervals chosen by \cite{Pageetal_2011} for spectral analysis are also shown by 
vertical lines (see text). Note that the final phase of the last pulse overlaps with the XRT observation 
from 125\,s. \emph{Lower panel:} Evolution of the hardness ratio (HR) of the BAT (25-50 keV)/(15-25 keV), 
and the XRT (1.5-10 keV)/(0.3-1.5 keV). We have fitted power-law functions to the piece-wise data (see text).
These are plotted by thick lines to guide the eye. The values of the flux and the hardness ratio are taken from 
{\tt http://www.swift.ac.uk/burst\_analyser/} (see \citealt{Evansetal_2007,Evansetal_2009}). 
}
\label{fig1}
\end{figure}

In Figure~\ref{fig1} (upper panel), we show the energy flux evolution (erg\,cm$^{-2}$\,s$^{-1}$) of 
GRB~090618 as detected by the BAT (in 15-50\,keV) and the XRT (in 0.3-10\,keV) detectors 
(filled and open symbols, respectively). The flux data is taken from UK Swift Science Data Centre. 
The time intervals used by \cite{Pageetal_2011} for the XRT analysis are shown by vertical solid lines.
\cite{Navaetal_2011}, fitting a Band function to the \emph{Fermi}/GBM data, calculates a time-integrated flux
in GBM band (over $\triangle t$=182.27\,s) to be $(3398\pm62)\times10^{-7}$ erg cm$^{-2}$, which is
one of the highest till date. The XRT data show a canonical flux evolution of the X-ray afterglow 
(e.g. \citealt{Zhang_2007_review}) i.e., a steep decay phase with a temporal index of $5.6\pm0.1$  
in 125-278\,s, followed by shallower decay phases with the index changing first to $1.9\pm0.2$ 
in 278-484\,s, and then to $0.67\pm0.02$ in 484-5150\,s (not shown in Figure~\ref{fig1}; 
see \citealt{Pageetal_2011} for details). 

During the prompt emission phase several broad pulses are seen. To study the average spectrum of the 
pulses, \cite{Raoetal_2011} have divided the lightcurve into four sectors representing the four 
pulses of this GRB. These sectors are 0-50\,s, 50-77\,s, 77-100\,s and 
100-180\,s, time being measured from the BAT trigger time. The spectrum in each sector is fitted with 
a Band function. The following parameters are obtained in these sectors (in sequence) --- peak energy 
($E_{\rm peak}$, in keV): $264_{-102}^{+209}$, $248_{-55}^{+81}$, $129_{-13}^{+19}$, and $33_{-14}^{+15}$,
low energy photon index ($\alpha$): $-1.18_{-0.08}^{+0.13}$, $-1.23\pm0.05$, $-1.39\pm0.03$, and 
$-1.70_{-0.10}^{+0.07}$, high energy photon index ($\beta$): $<-1.6$, $<-2.1$, $-2.4\pm0.2$, and 
$-2.8_{-1.1}^{+0.2}$. Note that the analysis indicates a spectral softening during the prompt 
emission. To study a detailed spectral evolution within each sector, \cite{Basak_Rao_2012_090618} have performed 
a pulse-wise analysis of this GRB. They have assumed an empirical function for the evolution of the peak energy 
(\citealt{LK_1996}): $E_{\rm peak}(t)=E_{\rm peak,0}~{\rm exp}\left[-\phi(t)/\phi_0\right]$,
where $E_{\rm peak}(t)$ is the peak energy at some observer time $t$, $E_{\rm peak,0}$ is the peak
energy at the beginning of the pulse, $\phi(t)$ is the time-integrated flux (or ``running fluence'') from 
the start of the pulse till $t$, and $\phi_0$ is a characteristic e-folding of the exponential evolution.
Note that this function signifies a hard-to-soft spectral evolution i.e., the peak energy starts with a high value 
($E_{\rm peak,0}$), and exponentially falls off with the ``running fluence''. As all the pulses could 
be described with such an evolution law, this indicates a spectral softening within the pulses of the GRB. 

Interestingly, \cite{Izzoetal_2012} have found evidence of two different episodes in the prompt emission 
data. In the first episode (0-50\,s) they have identified a thermal component which shows a temperature 
evolution resembling a hard-to-soft evolution. However, the signature of a thermal emission is insignificant in the 
rest of the burst. \cite{Basak_Rao_2013_parametrized}, using a ``parametrized joint fit'' in the rising and falling 
part of the first pulse (-1.0 to 40.85\,s), have indeed found evidence for a thermal emission. 
Interestingly, the thermal component consists of two correlated blackbodies rather than a single 
blackbody. The evidence is less clear in the subsequent pulses. Note in Figure~\ref{fig1} that the 
first pulse is well separated from the rest of the burst, while the intermediate pulses have large 
overlaps. Hence, it is possible that the true spectral evolution is disguised by the overlap of the 
pulses.  Note that the falling part of the last pulse (115\,s onwards) has a smooth flux decrement, and this part 
is relatively free from pulse overlap. Moreover, during this time, there is a simultaneous BAT and XRT 
observation, and \cite{Pageetal_2011} have found evidence of a blackbody component in the initial XRT data.
Hence, it is interesting to study the spectral evolution in the falling part of the last pulse.
The lower panel of Figure~\ref{fig1} shows the evolution of the hardness ratio (HR) for the BAT 
(black filled symbols) and the XRT (red open symbols) observations. We choose the following definition
of HR. BAT: (25-50 keV)/(15-25 keV), and XRT: (1.5-10 keV)/(0.3-1.5 keV), based on the criterion of
approximately equal observed counts in the respective energy bands. Note that the HR values show 
a hard-to-soft evolution in different sectors, namely, the first pulse (0-50\,s), falling part of 
the second pulse through the third pulse (60-100\,s) and the fourth pulse ($>100$\,s). The 
data in each of these sectors is fitted with a power-law function, which is over-plotted to guide 
the eye (solid curves). The HR data of the XRT observation is fitted in 125-205\,s interval. We note 
that the spectral softening of the last pulse is similar to the first two episodes. The XRT data in 
the overlapping region of the last pulse shows a similar evolution. Hence, the HR plot indicates that 
we are likely to get similar evolution of a thermal component in the BAT and the XRT spectral data during 
the falling part of the last pulse.

\section{Spectral Analysis and Interpretation}
\subsection{Comparison of Models in the Overlapping BAT-XRT Observation}
We first compare different models in the overlapping region of the BAT-XRT observation in the falling part of the 
last pulse. We perform a joint fit to the BAT and the XRT spectral data using a power-law, a Band, a BBPL,  
and a 2BBPL model. The $\chi^2$\,(and degrees of freedom --- dof) are 953.6 (285), 263.7 (283), 787.0 (283) 
and 307.1 (281), respectively. Clearly, the Band function gives a superior fit to the data.
However, for the Band function fit, we obtain a value of $\alpha=-0.71_{-0.08}^{+0.10}$. Note 
that this value is higher than the limit of synchrotron emission by $\sim9.9\sigma$ in the fast cooling 
regime, and by $\sim3\sigma$ in the moderately fast cooling regime (though compatible with 
the slow cooling regime). Hence, we conclude that though the Band 
function gives a statistically acceptable fit, we cannot associate this function with an optically thin 
synchrotron emission. Note that based on the $\chi^2$ value, the BBPL model is also not acceptable.
Although the 2BBPL has more free parameters than the Band function and results in 43 units 
of $\chi^2$ more than Band, we consider that this model may be physically more meaningful. 
Note that the power-law of the 2BBPL model has an index $\Gamma=-1.93\pm0.02$, well within 
the fast cooling synchrotron regime. That is, the power-law component of the 2BBPL model can
be associated with an optically thin synchrotron emission. Hence, we studied this model in more detail.
In addition, it is possible that the non-thermal component has different slopes in the lower 
and higher energies (see Section~\ref{PL}). Hence, the 2BBPL model can be an approximation of a more 
fundamental model. For example, a replacement of the power-law component of the 2BBPL model with a cut-off 
power-law gives $\chi^2$\,(dof)=250.4 (280). The power-law index is $-1.50_{-0.06}^{+0.05}$, which is again 
within the fast cooling synchrotron regime. 

\subsection{Evolution of the Thermal Component in the BAT and the XRT Data}

\begin{figure}\centering
{

\includegraphics[width=3.2in]{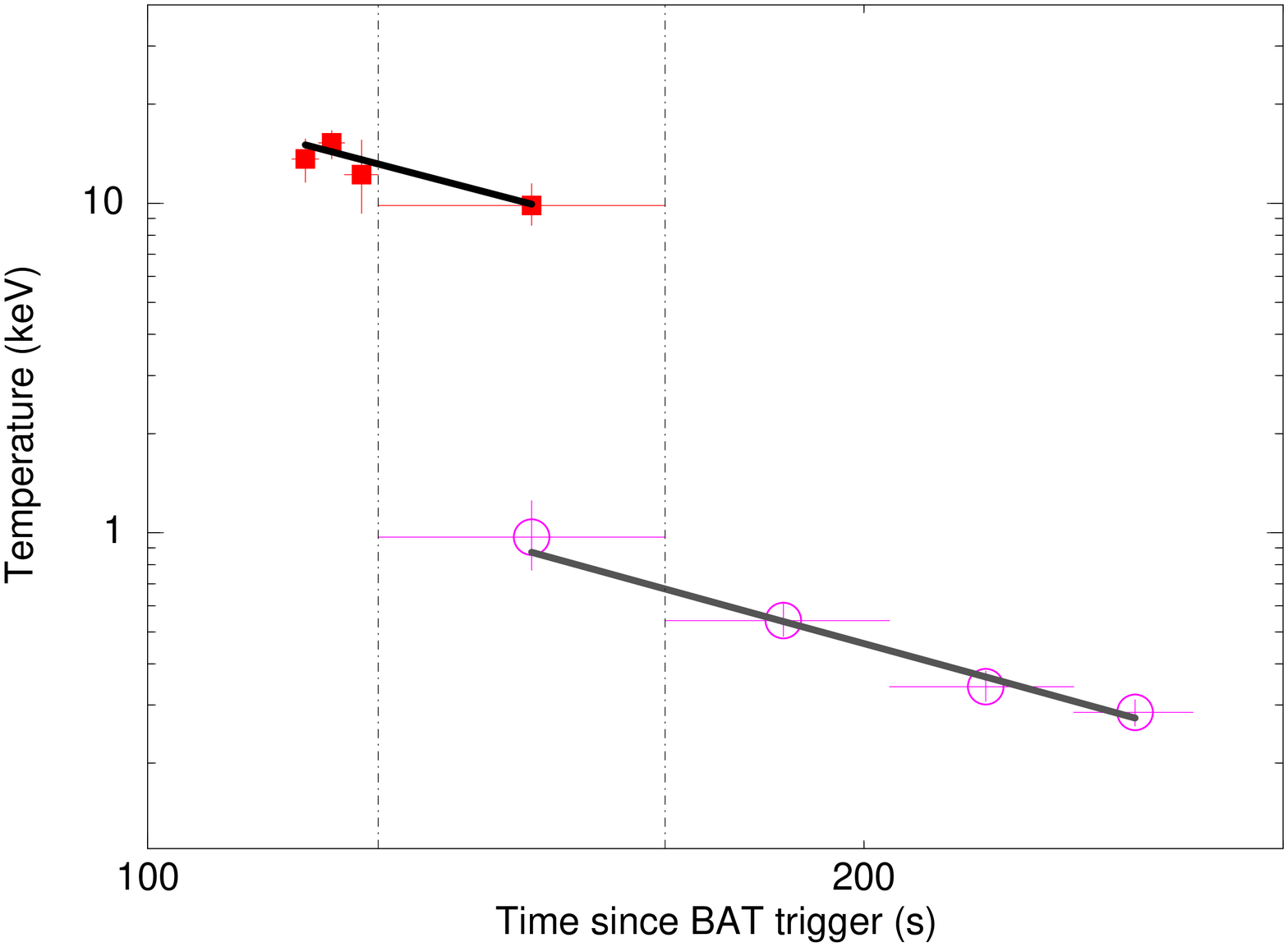} 

%\label{fig:Light Curves}
}
\caption{Temperature of the two blackbodies resulting from the BBPL fit to the BAT data (squares) 
and the XRT data (circles; from \citealt{Pageetal_2011}) as a function of time. The data in the 
overlapping region (125-165\,s) are  indicated by dotted vertical lines. We fit power-law function to the 
individual time evolution, and show them as straight lines with slopes $-1.9\pm0.9$ and $-2.0\pm0.5$ (at 68\% 
confidence level) for the higher- and lower-temperature blackbodies, respectively.
}
\label{fig2}
\end{figure}

\begin{table*}\centering
\caption{Parameters of spectral fitting to the time-resolved spectra of the final prompt emission 
phase of GRB 090618. We have used two models: (i) blackbody with a power-law (BBPL) and two blackbodies
with a power-law (2BBPL). During the initial time bins, the joint BAT-GBM data are fitted with a BBPL model. In 
the overlapping region of the BAT and XRT, a 2BBPL model is fitted. All times are measured from the time of the GBM trigger}
\vspace{0.3in}
\begin{tabular}{c|ccccccc}
\hline
Interval &  $kT$ or $kT_h$ & $N$ or $N_h$ & $kT_l$ & $N_l$ & $\Gamma$ & $\chi^2$ (dof)\\
\hline
\hline
\multicolumn{7}{l}{BAT-GBM data}\\
\hline
$116.95-118.05$ & $14.8_{-1.8}^{+1.8}$ & $1.7_{-0.5}^{+0.5}$ & --- & --- & $-2.2_{-0.1}^{+0.1}$ & 79.8 (104) \\

$118.05-119.35$ & $10.2_{-1.1}^{+1.4}$ & $1.4_{-0.3}^{+0.3}$ & --- & --- & $-2.1_{-0.1}^{+0.1}$ & 100.6 (104) \\

$119.35-120.95$ & $12.6_{-1.4}^{+1.5}$ & $1.4_{-0.3}^{+0.3}$ & --- & --- & $-2.2_{-0.1}^{+0.1}$ & 88.7 (104)\\

$120.95-122.65$ & $13.4_{-0.8}^{+0.8}$ & $2.1_{-0.3}^{+0.4}$ & --- & --- & $-2.4_{-0.1}^{+0.1}$ & 127.5 (104)  \\

$122.65-124.85$ & $14.6_{-2.7}^{+2.4}$ & $0.8_{-0.3}^{+0.3}$ & --- & --- & $-2.3_{-0.1}^{+0.1}$ & 123.0 (104)  \\

$124.85-127.25$ & $10.6_{-1.2}^{+1.4}$ & $0.9_{-0.4}^{+0.4}$ & --- & --- & $-2.3_{-0.1}^{+0.1}$ & 98.5 (104)  \\

$127.25-130.45$ & $8.4_{-1.3}^{+1.7}$ & $0.6_{-0.2}^{+0.2}$ & --- & --- & $-2.3_{-0.1}^{+0.1}$ & 93.3 (104)  \\
\hline
\multicolumn{7}{l}{BAT-XRT data}\\
\hline
$128.192-137.192$ & $7.54_{-0.32}^{+0.32}$ & $0.84_{-0.05}^{+0.05}$ & $1.24_{-0.10}^{+0.12}$ & $2.38_{-0.20}^{+0.23}$ & $-1.83_{-0.03}^{+0.03}$ &  165.71 (156) \\                    
$138.192-147.192$ & $7.00_{-0.46}^{+0.46}$ & $0.44_{-0.04}^{+0.04}$ & $0.99_{-0.07}^{+0.08}$ & $0.92_{-0.08}^{+0.08}$ & $-1.89_{-0.03}^{+0.03}$ &  151.85 (144) \\                    
$148.192-158.192$ & $5.87_{-1.39}^{+1.13}$ & $0.17_{-0.03}^{+0.03}$ & $0.91_{-0.08}^{+0.07}$ & $0.55_{-0.05}^{+0.05}$ & $-1.94_{-0.04}^{+0.03}$ &  179.50 (161) \\                    
$158.192-168.192$ & $4.82_{-0.82}^{+0.84}$ & $0.13_{-0.03}^{+0.03}$ & $0.68_{-0.04}^{+0.04}$ & $0.34_{-0.03}^{+0.03}$ & $-2.00_{-0.05}^{+0.04}$ &  136.78 (144) \\                    
\hline

\end{tabular}
\label{090618_a}
\vspace{0.3in} 
\begin{footnotesize}

\textbf{Notes:} Temperature ($kT$) and normalization ($N$) of a blackbody have usual units 
used in {\tt XSPEC}: keV and $10^{39}$\,erg\,s$^{-1}$\,(10 kpc)$^{-2}$.
Suffix `$h$' and `$l$' denote the higher and lower blackbodies, respectively. 
In the initial bins (where the XRT data is not available) the lower-temperature blackbody is speculated to be outside the
BAT energy band. Hence, the blackbody here represents the higher-temperature blackbody as detected in the higher energy detectors.\\

\end{footnotesize}

\end{table*}

\cite{Pageetal_2011} find evidence for an evolving blackbody component in the time-resolved spectra 
(125-165\,s, 165-205\,s, 205-245\,s, 245-275\,s and 275-2453\,s). 
In each case, the inclusion of a blackbody gives $>0.9999$ F-test significance over a power-law fit to 
the data. In Figure~\ref{fig2}, we have shown the blackbody temperature (open circles) in the first four 
intervals as used by \cite{Pageetal_2011}. The blackbody temperature steadily decreases 
(from $0.97_{-0.20}^{+0.28}$ keV to $0.285_{-0.026}^{+0.026}$ keV), clearly showing a cooling behaviour. 
From our analysis (\citealt{Basak_Rao_2013_parametrized}), we suspect that this blackbody is possibly the 
lower-temperature blackbody of the 2BBPL model. In order to find the signature of the higher-temperature blackbody, 
we use the BAT data, and perform a time-resolved spectroscopy using a BBPL model in the falling part of 
the last pulse. The time interval before the XRT observation ($<125$\,s) is divided into three nearly equal bins 
(115-118\,s, 118-121\,s, 121-125\,s). In addition, we use the first time bin used by \cite{Pageetal_2011} 
as one broad time bin (125-165\,s) to compare the temperature of the two blackbodies.
The temperature as obtained by our analysis is shown by filled squares 
in Figure~\ref{fig2}. The plot indicates a decrement of the temperature with time.
We fit a power-law function to the respective data points (the BAT and XRT). These are shown as solid lines
in the figure. Though there is a noticeable decreasing trend in the temperatures of both the data sets,  
the temperatures of the blackbodies during the overlapping observations are quite different in the BAT 
and the XRT data (differs by $>5\sigma$). Analysis of the data with the BBPL model (using the XRT data - 
\citealt{Pageetal_2011}) shows clear 
evidence for a black body component with temperature decreasing with time. 
In our analysis in the overlapping region (125-165\,s interval), the two different detectors (the BAT and XRT)
show distinctly different temperatures, indicating the possibility of having two black bodies as
possible spectral components. We examine the hypothesis that the underlying 
shape is two black bodies and a power-law and, in the following, we explore our hypothesis in detail.
Though  two peaks in the energy is required from the data, we note that alternative
interpretation of the two peaks is also possible (see \citealt{Guiriecetal_2015} who have used
a blackbody along with a  Band function and  a power-law).

\begin{figure}\centering
{

\includegraphics[width=3.2in]{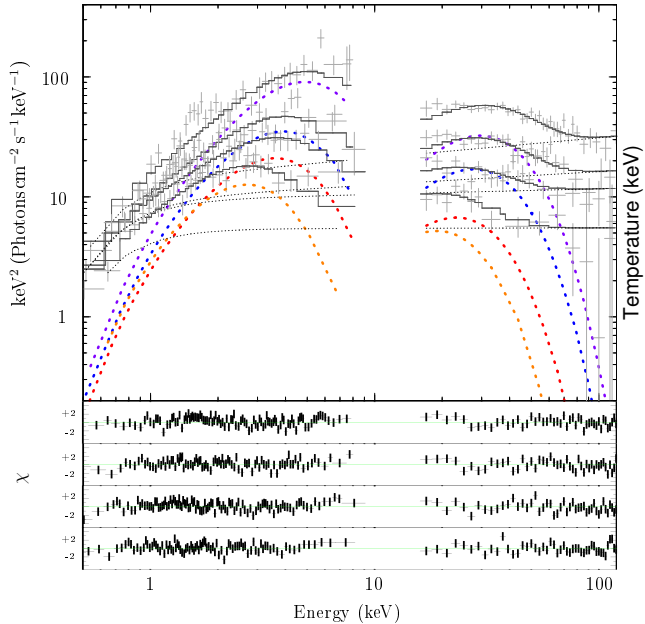} 

%\label{fig:Light Curves}
}
\caption{\textit{Top panel:} the data in the overlapping interval of the BAT and the XRT are fitted with a model consisting of two blackbodies
and a power-law (2BBPL). Shown are the spectral components of the model in the four time bins (see Table~\ref{090618_a}).
Time increases from top to bottom. The two blackbodies, the power-law, and the total model are shown by thick 
dotted curves, thin dotted curves, and histograms, respectively. Note that the temperature and normalization of each 
blackbody decrease with time. \textit{Lower panels:} residuals of the corresponding fits (lower panel for later time).
}
\label{fig3}
\end{figure}

Since there is a strong indication of spectral evolution in the first time bin of the BAT-XRT overlapping 
observation (see the HR plot in Figure~\ref{fig1}), we further subdivide the 125-165\,s data into four equal 
time bins. We use the 2BBPL model to fit the time-resolved spectra. In Figure~\ref{fig3}, we have shown the unfolded 
$\nu F\nu$ spectra with 2BBPL model fits (histograms) to the individual time-resolved data.
The two blackbodies in each time bin are shown by dotted curves. The observation time increases from the top 
to the bottom. The temperature and flux of both the blackbodies decrease with time. We note in Figure~\ref{fig3} 
that each detector captures one of the blackbodies of the 2BBPL model in the corresponding spectral window. Hence, in 
the absence of one of the detectors, the other `identifies' only a single blackbody. This must be happening before 125\,s 
when the XRT data is not available. An extrapolation of the temperature evolution, as seen in the XRT data, 
to 115\,s yields a temperature $\sim1.4$\,keV, well outside the lower energy coverage of the BAT. However, the 
lower-temperature blackbody can still be present in this data.
In order to find any signature of this blackbody during this time, we incorporate the GBM to perform a 
joint BAT-GBM analysis of the time-resolved data before 125\,s. Note that the lower energy coverage of the GBM extends 
down to $\sim 8$\,keV, hence, it can better identify any signature of the lower-temperature blackbody in the absence of 
the XRT observation. Following Section 2, we calculate the time intervals from the start of the pulse (100\,s).
We obtain 7 bins in the falling part of the pulse (116.95\,s to 130.45\,s). The 2BBPL model fitted to the joint 
BAT-GBM spectra gives the following $\chi^2$: 76.5, 95.5, 85.0, 123.8, 105.1, 91.3, and 91.5, with 102 dof in all cases.
However, the BBPL model also gave comparable fits with $\chi^2$: 79.8, 100.6, 88.7, 127.5, 123.0, 98.5, and 93.3, with 
104 dof in all cases. Hence, we can not significantly identify a low energy blackbody when fitting simultaneously 
the BAT and GBM data with the 2BBPL model. In Table~\ref{090618_a}, we have shown the best-fit values 
(with $1\sigma$ uncertainties) of the temperature ($kT$), normalization ($N$) and power-law index ($\Gamma$) of the
BBPL model. In this table, we also show the best-fit values of the 2BBPL model fitted to the data in the overlapping 
region of the BAT and the XRT observations. $kT_h$ and $N_h$ respectively denote the temperature and normalization of the
higher-temperature blackbody, while $kT_l$ and $N_l$ denote those of the lower-temperature blackbody.
All the time intervals are measured from the time of the GBM trigger. 
Note that the evolution of the temperature of the higher-temperature blackbody is consistent with that of the blackbody 
temperature in the previous bins. Hence, we conclude that the blackbody seen in the higher energy detectors is the 
higher-temperature blackbody of the 2BBPL model. It is worthwhile to mention that apart from 
analyzing the XRT data \cite{Pageetal_2011} also present a fitting of the joint BAT and GBM data.
In addition to the blackbody at around 0.2-1\,keV in the XRT data they find another blackbody 
at around 5-15\,keV, similar to our higher temperature blackbody (see their Figure 2, lower panel).
Hence, the two temperature solution is possible and not an artifact of the fitting technique we have developed.

\begin{figure}\centering
{
\hspace{-1in}
\includegraphics[width=3.2in]{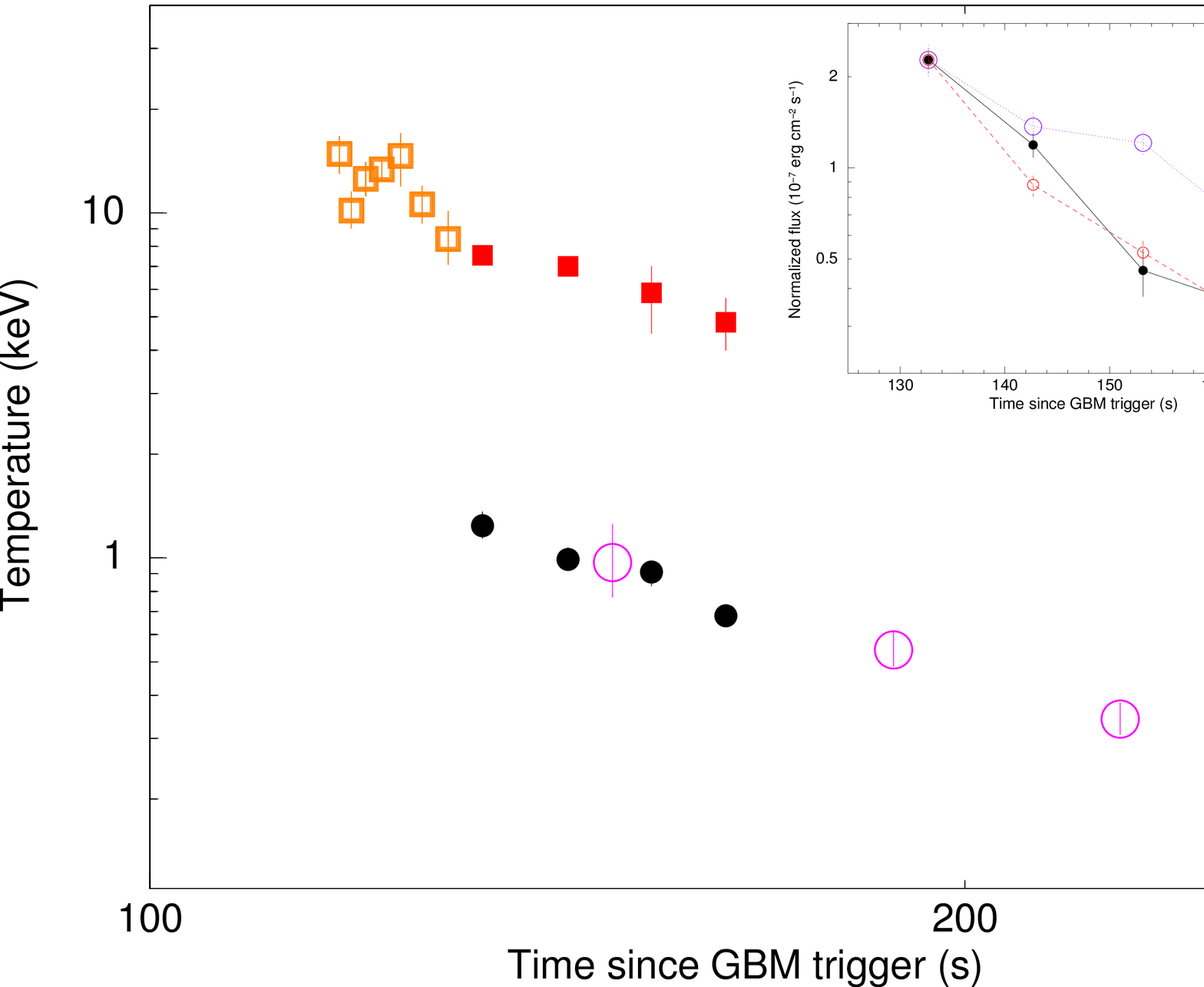} 

%\label{fig:Light Curves}
}
\caption{Evolution of the two blackbodies as obtained by various analyses. The open squares correspond to the 
temperature of the blackbody resulting from the BBPL fit to the joint BAT and GBM data. In the overlapping 
region of the BAT and the XRT observations, we have fitted the joint data with a 2BBPL model. Note that 
the evolution of the temperature of the higher-temperature blackbody (red filled squares) is consistent with the 
evolution of the blackbody temperature in the previous bins. The evolution of the temperature of the 
lower-temperature blackbody (filled circles) is consistent with the blackbody evolution in the XRT data 
(open circles, from \citealt{Pageetal_2011}). \emph{Inset:} The evolution of normalized energy flux of 
different components of the 2BBPL model during 125-165\,s  (see text for conversion factors) --- 
higher-temperature blackbody (filled circle), lower-temperature blackbody (small open circle), 
power-law (large open circle). Note that the flux of the thermal components rapidly falls off compared to the power-law.
}
\label{fig4}
\end{figure}

The temperature evolution as found in our analysis are graphically shown in Figure~\ref{fig4}. The squares represent 
the temperature of the higher-temperature blackbody, while the circles represent that of the the lower-temperature blackbody. 
The blackbody temperatures reported by \cite{Pageetal_2011} from the analysis of the XRT data are over-plotted (open circles). 
Note that the evolution of the blackbody temperature is consistent with that of the lower-temperature blackbody 
as obtained from the BAT-XRT joint analysis (filled circles). The evolution of the higher-temperature blackbody 
(filled squares) is also consistent with the evolution of the blackbody detected in the higher energy detectors before the 
XRT observation (open squares). We fit the time evolution of the blackbody temperatures with a power-law 
function. As the BAT data is used up to 165\,s, we restrict the fitting till this time. The indices of the power-law fit 
for the higher- and lower-temperature blackbody are $-3.6_{-0.5}^{+0.4}$ and $-2.8_{-0.5}^{+0.5}$, respectively, which are 
compatible with each other within $1.6\sigma$. Hence, there is a very strong indication that the two blackbodies detected in the lower and 
higher energy detectors are two components of a unified 2BBPL model. Both these components show a smooth evolution during 
the junction of the late prompt emission and the early afterglow phase. 

Another important conclusion can be drawn by comparing the flux evolution of each component of the 2BBPL 
model during 125-165\,s. 
This interval is chosen because there is a simultaneous observation of all the components 
of the 2BBPL model in this 
interval. Note that the lightcurve shows a steep decay during this time. 
The steep decay phase can be a combination of high 
latitude emission due to ``curvature effect'', and the canonical afterglow 
decay (\citealt{Zhang_2007_review}). The first 
effect leads to a rapid flux decrement, while the second effect is likely to show a slower decay.
\cite{Pageetal_2011} have shown in the XRT data that the contribution of the 
blackbody to the total flux decreases with time, 
and the signature of a thermal emission is absent in the late XRT and 
\emph{XMM-Newton} data. In the inset of Figure~\ref{fig4}, 
we show the evolution of the flux ($10^{-7}$\,erg\,cm$^{-2}$\,s$^{-1}$ 
in 0.1-150\,keV) of each component of the 2BBPL model. 
In order to show the flux evolution of the components in the same scale, 
the flux of the blackbodies are normalized 
with the power-law flux at the first bin. 
The multiplication factor to get the actual flux values are 
0.31 and 0.94 for the higher-temperature blackbody and the lower 
blackbody, respectively. It is interesting to note that the 
two blackbodies show a steeper decay than the power-law 
component. This is consistent with the findings of 
\cite{Pageetal_2011}. Hence, the thermal emission is probably the 
late phase of the prompt emission. 

\subsection{Evolution of the Photosphere}

\begin{figure}\centering
{

\includegraphics[width=3.2in]{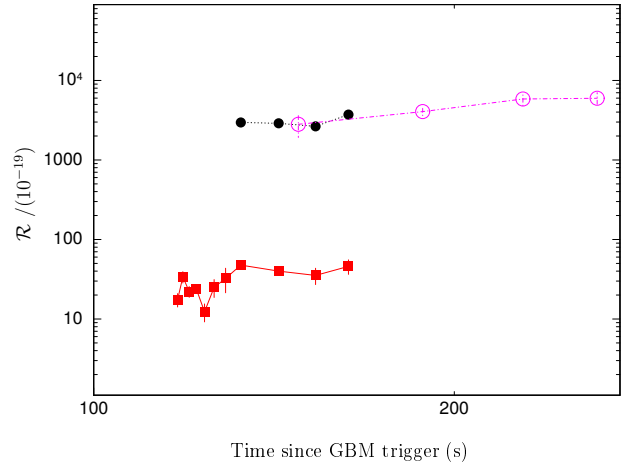} 

%\label{fig:Light Curves}
}
\caption{The evolution of $\cal{R}$/($10^{-19}$) for the higher-temperature blackbody (filled squares joined by 
continuous line), and the lower-temperature blackbody (filled circles joined by dotted line) are shown. The values 
of $\cal{R}$ calculated from \cite{Pageetal_2011} are over-plotted. The physical photospheric radius is directly 
proportional to $\cal{R}$ (\citealt{Ryde_Pe'er_2009}). Note that the photosphere corresponding to the 
lower-temperature blackbody remains at a higher radius compared to that of the higher-temperature blackbody. 
Both the photospheres show a very slow and overall increase in size.
}
\label{fig5}
\end{figure}

In the fireball model of GRBs (e.g., \citealt{Goodman_1986}), thermal emission is expected from a photosphere
where the optical depth is $\sim 1$, and the radiation decouples from the matter. If the two blackbody emissions 
found in our analysis are related to two such photospheres, it is interesting to study the evolution of the 
photospheric radius ($r_{\rm ph}$) during the transition of the prompt and afterglow phase. As the temperature 
and flux vary smoothly during this time, we expect a smooth photospheric evolution. To compute $r_{\rm ph}$, 
let us define a dimensionless quantity ${\cal{R}}$ (see \citealt{Ryde_Pe'er_2009}) as follows.
 
\begin{equation}
 {\cal{R}} (t)=\left[\frac{F_{\rm BB}(t)}{\sigma T(t)^4} \right]^{1/2}
\label{R}
\end{equation}
\vspace{0.1in}

where $F_{\rm BB} (t)$ is the energy flux in units of erg\,cm$^{-2}$\,s$^{-1}$ and $T(t)$ is the 
temperature of a blackbody (in units of Kelvin, K) at an observer time, $t$. $\sigma$ is Stefan-Boltzmann 
constant $=5.6704\times10^{-5}$ erg\,cm$^{-2}$\,s$^{-1}$\,K$^{-4}$. As the photospheric radius is 
directly proportional to ${\cal{R}}$, it is sufficient to study the evolution of ${\cal{R}}$. We 
calculate the flux of the individual blackbodies and calculate the corresponding ${\cal{R}}$. 
The uncertainty in ${\cal{R}}$ is calculated by propagating the uncertainties of temperature and flux. In 
Figure~\ref{fig5}, we show the time evolution of ${\cal{R}}$. We note that (i) the evolutions are 
similar in the overlapping region, (ii) the photosphere corresponding to the lower-temperature blackbody 
is at a higher radius than the photosphere corresponding to the higher-temperature blackbody, and 
(iii) each photosphere shows an overall increase in size.

\begin{figure}\centering
{

\includegraphics[width=3.2in]{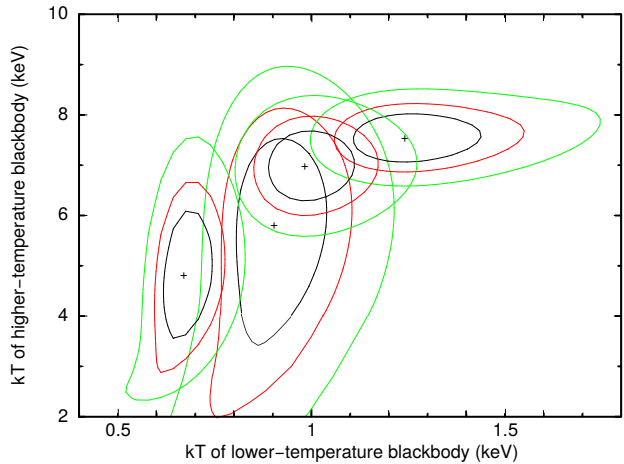} 

%\label{fig:Light Curves}
}
\caption{$\chi^2$ contour plot for the temperature of the higher-temperature blackbody ($kT_h$) and the temperature
of the lower-temperature ($kT_l$) blackbody. A plus symbol denotes a pair of ($kT_h$, $kT_l$) value obtained by $\chi^2$ 
minimization for one time interval. We have shown three contour levels (68\%, 90\% and 99\%) for each case.
If the correlation between the two parameters are due to mathematics, then we expect elliptical contours with major axis 
along $45^{\circ}$ angle. The fact that the contours are either nearly circular or elliptical with major axis parallel to 
either x- or y-axis shows that the correlation between $kT_h$ and $kT_l$ are not due to mathematics.
}
\label{fig4a}
\end{figure}

From Table~\ref{090618_a}, we note that the temperature and normalization of the two 
blackbodies are highly correlated in the simultaneous BAT and XRT data. The temperature of the 
two blackbodies have a Pearson correlation coefficient, $r=0.96$ with a chance probability, 
$P_r=4\times10^{-2}$. The normalization of the two blackbodies have $r=0.98$, and $P_r=2\times10^{-2}$. 
In order to investigate whether this correlation indicates a real physical picture or a mathematical 
artifact, we plot the $\chi^2$ contour for $kT_h$ and $kT_l$ (see Figure~\ref{fig4a}). We note that 
none of the contours have elliptical shape with major axis along $45^{\circ}$ angle, which is expected 
if the parameters co vary. Hence, we conclude that the correlation is both significant and real.
Since the photospheric radius is calculated using the temperature and the normalization, it is natural 
that the photospheres corresponding to the two blackbodies have similar evolution. To investigate this 
further, we tie the ratio of the temperature and normalization of the blackbodies in all bins 
of the simultaneous BAT and XRT data (see Section 2). In addition, 
as we have found little variation of the power-law index ($\Gamma$), we tie this parameter in all 
time bins to reduce the number of free parameters. With these constraints, 
we obtain $\chi^2_{\rm red}$ (dof)=1.05 (607). With $\Gamma$ as 
free parameter we obtain $\chi^2_{\rm red}$\,(dof)=1.03 (604), and with no constraints, 
$\chi^2_{\rm red}$ (dof)=1.03 (598). Note that while we get similar $\chi^2_{\rm red}$, 
the `dof' increase with tied fitting. Moreover, such tied fitting confirms that the two blackbodies 
are indeed correlated physically and the result is not due to mathematics. We obtain the ratio
of temperature and normalization of the higher-temperature blackbody 
to the lower-temperature blackbody as $6.40\pm0.39$ and $0.40\pm0.03$, respectively.
As the normalization of the blackbodies are given in energy units, the energy flux ratio of the 
blackbodies is equal to the normalization ratio. Hence, putting these values in Equation~\ref{R}, we obtain the 
ratio of $\cal{R}$ as

\begin{equation}
\frac{\cal{R}_{\rm low}}{\cal{R}_{\rm high}}=\left(\frac{N_{l}}{N_{h}}.\frac{T_{h}^4}{T_{l}^4} \right)^{1/2}\sim 65\pm8.3
\end{equation}
\vspace{0.1in}

Hence, the value of $\cal{R}$ corresponding to the lower-temperature blackbody is about 65 times larger than that of
the higher-temperature blackbody (see Figure~\ref{fig5}). 

\subsection{The Spectral Shape of the Non-thermal Component}\label{PL}
We have seen that the thermal components of the 2BBPL have smooth evolution during the transition of the prompt 
emission to the early afterglow phase. Let us now investigate the evolution of the power-law component. 
\cite{Pageetal_2011} have found 
the following indices ($\Gamma$) in the four time bins: $-1.23_{-0.19}^{+0.20}$, $-1.77_{-0.12}^{+0.10}$, 
$-1.84_{-0.13}^{+0.12}$, and $-1.85_{-0.14}^{+0.12}$. As the flux rapidly evolves in the first time 
bin, it is better to use finer bins, as we have done in our analysis. Note from Table~\ref{090618_a}
that the values of $\Gamma$ before the XRT observation (i.e., $<125$\,s) are much lower than these values. 
There can be two reasons for this observation --- (i) the index evolves, or (ii) there is a difference in the 
spectral index at lower and higher energies. 

In order to investigate this, we perform a tied fitting (see Section 2) of the BAT-GBM data before the 
XRT observation. We incorporate a lower-temperature blackbody with the ratio of temperature and norm tied 
to those of the higher-temperature blackbody. We use the ratios as obtained by the tied fitting of the 
BAT-XRT joint data (see Section 4.3). In addition, the temperature of the lower-temperature blackbody 
is restricted below 5\,keV (by putting an upper bound to the parameter), and the parameter 
$\Gamma$ is tied in all bins. With these assumptions, we obtain $\Gamma=-2.16_{-0.04}^{+0.05}$.
Note that the value of $\Gamma$ is comparable with those obtained by fitting the individual time-resolved 
spectra of the BAT and GBM joint data. However, this value is quite different from the $\Gamma$ values
(in the range $-1.23\pm0.20$ to $-1.85\pm0.12$) resulting from the BBPL fitting to
the XRT data (\citealt{Pageetal_2011}). An evolution in $\Gamma$ cannot account for this drastic change. 
Hence, the difference possibly signifies a difference in the slope of the spectrum 
in the lower and higher energies, and a steeper slope is naturally obtained in the 
absence of the low energy data. Note the values of $\Gamma$ in the BAT-XRT joint 
data are indeed intermediate to these values (see Table~\ref{090618_a}). Hence, the non-thermal component of 
the spectrum probably has a steeper slope at higher energies. To investigate this, we perform a tied fitting 
of the BAT-XRT joint data by replacing the power-law component of the 2BBPL model with a cut-off power-law (CPL)
function. The CPL spectral model can be written as

\begin{equation}
 F(E)=K~E^{-\alpha_{\rm CPL}}{\rm exp}\left(-\frac{E}{\beta_{\rm CPL}}\right)
\end{equation}

where $\alpha_{\rm CPL}$ is the power-law index, $\beta_{\rm CPL}$ is the e-folding energy of the exponential rolloff 
(in keV), and $K$ is the norm in Photons\,keV$^{-1}$\,cm$^{-2}$\,s$^{-1}$ at 1 keV.
We tie the values of $\alpha_{\rm CPL}$ and $\beta_{\rm CPL}$ in all time bins. The ratio and normalization of 
the two blackbodies are fixed to those obtained in the tied analysis. We obtain $\chi_{\rm red}^2$\,(dof)=0.93\,(606).
Note that compared to the tied 2BBPL model fit ($\chi_{\rm red}^2$\,(dof)=1.05\,(607)), this is marginally better.
We obtain the following parameters of the CPL component: $\alpha_{\rm CPL}=-1.78\pm0.03$ 
and $\beta_{\rm CPL}=169.8_{-33.6}^{+54.9}$\,keV. Now, the index in the energy range $E_1$ to $E_2$ 
can be approximated by $\frac{{\rm log_{10}[F(E_2)/F(E_1)]}}{{\rm log[E_2/E_1]}}$.
The slope in the XRT energy band ($0.3-10$\,keV) is found to be $-1.80\pm0.03$, while the slope in the BAT energy band 
($10-150$\,keV) is found to be $-2.12\pm0.12$. Note that these values are remarkably close to those obtained by 
independent fitting of the XRT and the BAT data, respectively. This analysis illustrates that the slope of the non-thermal 
component has indeed a spectral variation rather than a time evolution.

% we replace the power-law function with 
% a cut-off power-law, and fit the BAT-XRT joint data. The ratio and normalization of the two blackbodies are fixed to 
% those obtained in the tied analysis. For example, in the first bin, the non-thermal component has the following 
% functional form: $F(E)\sim E^{-1.74\pm0.05} {\rm exp}\left(\frac{E}{223_{-71}^{+199}}\right)$. Now, the
% index in the energy range $E_1$ to $E_2$ can be approximated by $\frac{{\rm log_{10}[F(E_2)/F(E_1)]}}{{\rm log[E_2/E_1]}}$.
% The slope in the XRT energy band ($0.3-10$\,keV) is found to be $-1.75\pm0.05$, while the slope in the BAT energy band 
% ($10-150$\,keV) is found to be $-2.00\pm0.18$. A slope in the GBM/NaI energy band ($8-900$\,keV) is $-2.58\pm0.44$.
% This analysis illustrates that the slope of the non-thermal component has indeed a spectral variation rather 
% than a time evolution.

\section{A Physical Model For The Origin of 2BBPL}

\subsection{A Spine-sheath Jet Model}

\begin{figure*}\centering
{

\includegraphics[width=5.9in]{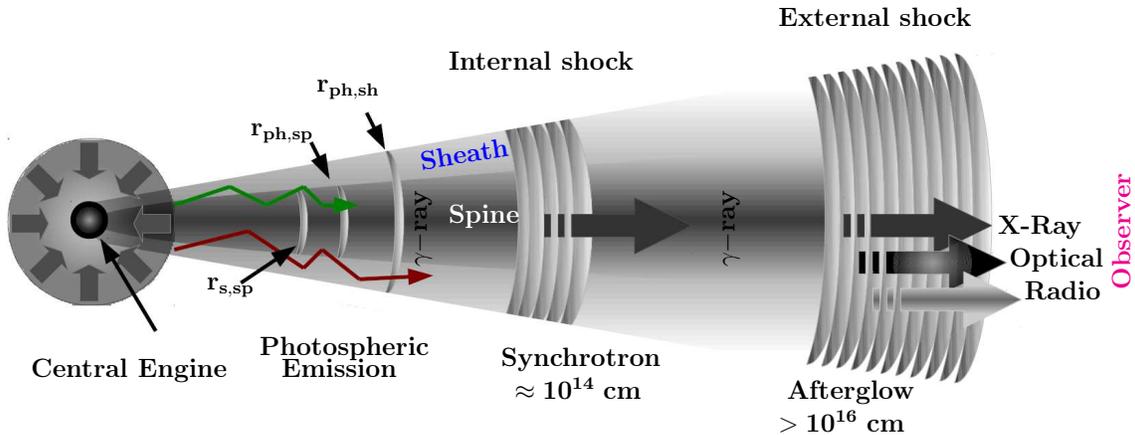} 

%\label{fig:Light Curves}
}
\caption{A schematic view of the spine-sheath jet model. This model naturally explains the presence of 
two blackbodies in the spectrum. In addition, such a jet can generate a cut-off power-law by inverse-Comptonizing 
the photons which cross the spine-sheath boundary (red and green zig-zag path). We have marked the location of 
the photospheres of the individual components --- $r_{ph, sp}$ for the spine and $r_{ph, sp}$ for the sheath, 
respectively. Note that due higher baryon loading, the photosphere of the sheath occurs at a higher radius.
$r_{s, sp}$ denotes the location of the saturation radius of the spine, which may be lower or higher than $r_{ph, sp}$.
We also mark the internal/external shock regions.
}
\label{fig6}
\end{figure*}

The model we propose to explain all the observations regarding the 2BBPL model is a spine-sheath jet model
(see Figure~\ref{fig6}). This is one of the commonly discussed jet structure in the literature
(e.g., \citealt{Meszaros_Rees_2001, Ramirez-Ruizetal_2002, Vlahakisetal_2003}). From the theoretical point of 
view, such a structure is expected as the GRB jet moves through the envelope of the progenitor star. The 
material of the progenitor forms a hot cocoon on the fast moving jet (\citealt{Meszaros_Rees_2001, 
Ramirez-Ruizetal_2002, Zhangetal_2003, Zhangetal_2004}).
Alternatively, a collimated proton jet surrounded by a less collimated neutron sheath is naturally 
expected for a jet with considerable magnetic field (\citealt{Vlahakisetal_2003, Pengetal_2005}). On the observational side, a spine-sheath
jet structure is frequently invoked to explain various observations e.g., shallow decay phase during the X-ray 
afterglow (\citealt{Granotetal_2006, Panaitescuetal_2006, Jinetal_2007, Panaitescu_2007}), double jet 
break and optical re-brightening (\citealt{Bergeretal_2003_spinesheath, Liang_Dai_2004, dePasqualeetal_2009, Hollandetal_2012}). 
For example, \cite{Bergeretal_2003_spinesheath}, using the afterglow data of GRB 030329 have found evidence 
of a collimated spine with an opening angle,
$\theta_{\rm spine}\sim 5^{\circ}$, and a wider sheath with an opening angle, $\theta_{\rm sheath}\sim 17^{\circ}$.
Note that for such a large opening angle, the thermal component of the sheath should have a shape of
a multicolour blackbody.
Interestingly, the collimation corrected energy of the sheath is 5 times higher than that of the spine.
Hence, most of the jet power is stored in the sheath component.
On the other hand, \cite{Hollandetal_2012}, using the optical afterglow data of GRB 030329, have found 
a much tighter collimated sheath. The opening angle of the spine and sheath are:  $\theta_{\rm spine}\sim 0.86^{\circ}$, 
and $\theta_{\rm sheath}\sim 1.4^{\circ}$. In addition, the jet luminosity of the two components 
are comparable with each other. Hence, a variety of spine-sheath structure is possible. 
Note that if we consider high latitude emission, the spectrum is expected to be broadened. 
In the spine-sheath model, the high latitude emission is produced by the sheath. Hence, the blackbody 
spectrum from the sheath can be broadened. However, the fact that the spectrum at lower energies can 
be adequately fitted with a blackbody indeed signifies that the opening angle of the sheath is possibly 
not very wide (in accordance with \citealt{Hollandetal_2012}).

Recently, \citet[][I13 hereafter]{Itoetal_2013} have performed a \emph{Monte Carlo} simulation of a spine-sheath 
jet. They consider $0.5^{\circ}$ and $1^{\circ}$ of opening angle for the two components, respectively. They vary 
the ratio of the speed and viewing angle, and obtain a variety of synthetic spectra (Figure 5 of I13). Interestingly,
they find an apparent signature of the two blackbodies for a viewing angle near the spine-sheath boundary. In addition,
they find a cut-off power-law component with a cut-off at $\sim10$\,MeV (H. Ito, private communications).  

\subsection{Explanation of the Observations}
Following I13, we consider the spectral evolution in such a spine sheath jet.
First, let us define the required quantities. The coasting bulk \emph{Lorentz} factor
is defined as $\eta=L/\dot{M}c^2$, where $L$ is jet luminosity, and $\dot{M}$ is the mass flow rate. The photospheric
radius, $r_{\rm ph} \propto L\eta^{-3}$. The radius where the bulk \emph{Lorentz} factor ($\Gamma_{\rm B}$) saturates 
is called the saturation radius, $r_{\rm s} \propto r_{\rm i}\eta$. Here, $r_{\rm i}$ is the position where the initial 
energy is injected. The quantities related to spine and sheath are denoted by the suffix `sp' and `sh', respectively. 
We assume $\eta_{\rm sp}>\eta_{\rm sh}$. Let us now estimate the parameters from the observations. 

(i) Note that the model naturally explains the presence of two correlated blackbodies. The evolution of 
$\Gamma_{\rm B}$ of the individual components is $\Gamma_{\rm B}\propto r/r_{\rm i}$ till $r_{\rm s, sh}$. 
Beyond this radius, while the sheath coasts with $\Gamma_{\rm B,sh}=\eta_{\rm sh}$, the spine continues to 
accelerate till $r_{\rm s, sp}$. The temperature in the comoving frame, $kT'$ has the following evolution (see I13).

\begin{equation}
 kT'\propto \left(\frac{L}{r_{\rm i}^2}\right)^{1/4}\times \left\{ \begin{array}{ll}
 (r/r_{\rm i})^{-1} & \mbox{$r < r_{\rm s}$} \\
 (r_{\rm s}/r_{\rm i})^{-1} (r/r_{\rm s})^{-2/3} &\mbox{$r > r_{\rm s}$}
       \end{array} \right.
\label{i13_1}
\end{equation}
\vspace{0.1in}

where $r$ is the radial distance measured from the centre of explosion in the lab frame. As the 
observed temperature, $kT\sim \Gamma_{\rm B} kT'$, the temperature remains constant till the saturation radius,
and then falls off as $r^{-2/3}$. Now, in the falling part of a pulse, we have observed that the temperatures 
of both the blackbodies monotonically decreases. Hence, the photosphere must have occurred above the saturation
radius for both spine and sheath. From Equation~\ref{i13_1}, the observed temperature at $r_{\rm ph}$ is 
$kT(r_{\rm ph}) \propto r_{\rm i}^{1/6} \eta^{8/3} L^{-5/12}$. Hence, the temperatures at the respective photospheres
strongly depend on the value of $\eta$.  Also, as $\eta_{\rm sp}>\eta_{\rm sh}$, the spine photosphere ($r_{\rm ph,sp}$) 
occurs below the sheath photosphere ($r_{\rm ph,sh}$). We can calculate the required ratio of $\eta$ from the observed 
ratio of $\cal{R}$ ($\sim65$; see Equation 2).

\begin{equation}
 \frac{\eta_{\rm sp}}{\eta_{\rm sh}} \sim \left(\frac{L_{\rm sp}}{L_{\rm sh}}.\frac{r_{\rm ph, sh}}{r_{\rm ph, sp}}\right)^{1/3}
\end{equation}

If we assume that $L$ scales with $\eta$ (as considered by I13), then we find a ratio of $\frac{\eta_{\rm sp}}{\eta_{\rm sh}} \sim 8$.
If, on the other hand, the spine and sheath have similar values of $L$, then the ratio is $\sim 4$. If we also assume that 
$r=\eta \cal{R}$, then a ratio of 2-4 is obtained.

(ii) Let us explain the observation regarding the non-thermal component of the spectrum.
For this, we shall use the results of \cite{Basak_Rao_2013_linger}. They have shown that 
the non-thermal component of the 2BBPL model 
has a strong connection with the GeV emission of a GRB. Phenomenologically, one can find 
two very different classes of GRBs with GeV emission --- hyper-fluent LAT GRBs (e.g., 
GRB~0909002B, GRB~090926A), and low-fluent LAT GRBs (e.g. GRB 100724B). Though these two 
classes have similar keV-MeV brightness, the hyper-fluent LAT GRBs have an order of magnitude 
higher GeV emission than the other class of GRBs. \cite{Basak_Rao_2013_linger} have 
shown that the power-law component of a hyper-fluent LAT GRB has a delayed onset and a 
lingering behaviour with respect to the thermal component. However, the low-fluent LAT GRBs 
show a coupled thermal and non-thermal variability. Let us explain this behaviour in the framework 
of the structured jet model.

First, we note that at a radius higher than $r_{\rm s, sh}$, the bulk {\emph Lorentz} factor of the 
sheath coasts with a constant value while that of the spine still accelerates.
Hence, a velocity shear occurs between these two regions above this radius. 
The photons which cross the spine-sheath boundary in this region 
effectively gains energy by inverse-Compton effect, and forms a power-law distribution. However, 
as the photon energy increases, the Compton scattering approaches the \emph{Klein-Nishina}
regime and the scattering cross section rapidly decreases. Hence, we expect a cut-off in the
emergent power-law component. I13 obtain a cut-off at $\sim 100 $ MeV, corresponding to
a $\Gamma \sim 200$. In addition, internal shock above the photosphere can lead to 
synchrotron emission. The synchrotron radiation above the photosphere is delayed compared to the thermal emission. 
Hence, for the hyper-fluent LAT GRBs, this is possibly the dominant process for the production of the non-thermal 
component. As the radiation is expected to have a delayed onset and a longer lasting behaviour compared 
to the thermal emission, this model is consistent with our findings. The other mechanism is nearly simultaneous 
with the photospheric emission. Hence, this process is likely to be dominant for the low-fluent LAT GRBs which 
show a simultaneous evolution of the thermal and the non-thermal component.
 
In Figure~\ref{fig6}, we have shown a schematic picture of the spine-sheath jet model.
Each of the components has its individual photosphere. As the spine has higher $\eta$, the 
photospheric radius ($r_{\rm ph, sp}$) is lower than that of the sheath
($r_{\rm ph, sh}$). The photons crossing the spine-sheath boundary layer effectively form a power-law with 
a cut-off. In addition, internal shocks above the photosphere also contributes to the non-thermal emission.
All these emissions constitute the prompt emission phase. Note that the emissions from the spine and sheath 
components can be seen both during the prompt and the afterglow phases.
 
\section{Discussion  and Conclusions}
In recent works, we have advocated the 2BBPL model as the universal model of GRB prompt emission data 
(\citealt{Basak_Rao_2013_parametrized, Basak_Rao_2013_linger, Basak_Rao_2014_MNRAS, Raoetal_2014}). We have 
studied the \emph{Fermi}/GBM data of various categories of GRBs --- single pulse, multiple separable pulses, and 
multiple overlapping pulses. In all cases, we have found that the 2BBPL model is consistent with the data.
As a single model can be applied for all categories of GRBs, it strongly indicates a common radiation mechanism. 
%The current analysis indeed shows the existence of the two blackbodies in the prompt emission spectrum and  
%as the spectra evolve (with temperature decreasing with time) the lower black body component is
%captured by the superior lower energy detectors.
The current analysis puts the blackbody component seen in the XRT data (\citealt{Pageetal_2011}) in the context of the  comprehensive 
2BBPL model:   two blackbodies are seen in the prompt emission spectrum and  
as the spectra evolve (with temperature decreasing with time) the lower black body component is
captured by the superior lower energy detectors.
In order to explain all the observations related to the 2BBPL model, we have proposed a physical 
picture, namely, a spine-sheath structure of a GRB jet. This model could explain the existence
of the two blackbodies and the relative distance of the corresponding photosphere. The evolution of the 
non-thermal component and its connection with the GeV emission is also explained.

It is worthwhile to mention that though the 2BBPL model has been observed in several GRBs, the spine-sheath 
model may not be the only possible physical setting. It is possible that the model considered here is a 
simplification of a more physical and complex model. A modification of the model, or a completely different 
process cannot be ruled out. For example, in the Cannonball model (\citealt{Dadoetal_2002,
Dar_Rujula_2004}), the central engine releases a sequence of ordinary matter, called 
``cannon-balls'' (CB). The energy is liberated via particle interaction rather than a photospheric
emission or shock generation. The CBs give rise to a blue-shifted bremsstraulung spectrum and 
inverse Comptonize the ambient photons. The higher-temperature blackbody of the 2BBPL model can be identified 
with the photons boosted by the interaction of CBs with the supernova shell, and the lower-temperature blackbody 
is possibly associated with the bremsstrahlung photons. As the two components are produced by the same 
CB, they should be correlated. Even in the framework of the fireball   model, dissipative processes 
(e.g., magnetic reconnection, or internal shocks) can re-energize the thermal emission by Comptonization
(\citealt{rees_and_meszaros_2005_prompt}). For fast dissipation, the pairs generated in the process can 
form an effective pair photosphere. Hence, the thermal emission from both the spine and sheath can be 
Comptonized blackbody emission (``grey body'') from the respective pair photospheres. A variety of such 
modifications over the simple two-component model, in principle, can be conceived.

In addition, the origin of the spine and sheath components of a GRB jet is debatable. 
\cite{Starlingetal_2012}, e.g., have shown for a few GRBs including GRB 090618
that the thermal emission possibly comes from the cocoon and the high flux associated with
these GRBs may challenge the scenario of thermal emission due to supernova 
shock breakout (\citealt{Campanaetal_2006, Waxmanetal_2007, Nakar_Sari_2010}).
In addition, our analysis shows that the photospheric radius of each component
increases very slowly, while the temperature falls off quite rapidly (see Figure~\ref{fig4}
inset and Figure~\ref{fig4a}). This result agrees with that obtained by \cite{Pageetal_2011} 
and \cite{Starlingetal_2012}, but disagrees with the shock breakout scenario in which the radius 
should increase as $\sim t^{0.8}$ and the temperature should decrease as $t^{-0.5}$.
If the sheath is the cocoon, the opening angle should be wide ($\sim10^{\circ}-20^{\circ}$), and the value 
of $\eta$ should be about 5-10 times lower than the spine (e.g., \citealt{Zhangetal_2004}). Note that 
in our analysis, we require a ratio of $\eta$ in this range. However, the ratio crucially depends on the 
jet luminosity ($L$). The observation of the prompt emission alone cannot determine all the 
parameters of the two components (e.g., $L$, opening angles etc). In this regard, the afterglow 
observation can give further clue on the structure of a GRB jet, provided the two jet breaks are 
properly identified. Note that the afterglow observation of \cite{Bergeretal_2003_spinesheath} and 
\cite{Hollandetal_2012} require very different components of a structured jet in terms of the opening 
angle and energetics. Hence, a variety of jet structures, in principle, can exist.

To summarize, motivated by the detection of a black body component with 
temperature decreasing with time in the early afterglow of GRB 090618 
(\citealt{Pageetal_2011}), we investigate the spectral evolution from the
peak of the last pulse in the prompt emission (T$_0$+117\,s; T$_{90}$ = 113\,s)
to the start of the shallow decay phase of the afterglow emission (T$_0$+275\,s),
using data from the \emph{Swift}/XRT, the \emph{Swift}/BAT and the \emph{Fermi}/GBM detectors.
The major findings of this work are as follows:

1. The light curve and the hardness ratio (HR) plots indicate that 
throughout this late prompt/ early afterglow phase, the spectral evolution
is hard-to-soft, strongly resembling the HR behaviour during the prominent
pulses of the prompt emission.

2. When we fit the data with a model consisting of a blackbody and a power-law
(the BBPL model), as was done by \cite{Pageetal_2011} for the \emph{Swift}/XRT data, we notice
a distinct trend of temperature decreasing with time. Remarkably, the temperature
obtained for the hard X-ray data (from the \emph{Swift}/BAT) is a factor of 6 higher than that 
obtained in the low energy data (from the \emph{Swift}/XRT).

3. We fit the simultaneous BAT and XRT data with various models. By comparing the $\chi^2$ of the 
fits, we find that the Band function is the best model. The next acceptable model is a model consisting 
of two blackbodies and a power-law (2BBPL). Though the Band function is statistically the best, the model is 
not physically motivated, and further the low energy spectral index of the model is found to be 
inconsistent with a fast cooling or moderately fast cooling regime of synchrotron radiation. 
Hence, we further investigate the 2BBPL model. We have also shown that replacing the power-law of 
the 2BBPL model with a cut-off power-law gives a marginally better fit than the Band function
but with three more free parameters.

4. By employing various statistical techniques like tied spectral fitting, we find that 
the data is consistent with a 2BBPL model. The ratio of temperatures of the
two black bodies is found to be 6.40$\pm$0.39 and the ratio of normalization is
found to be 0.40$\pm$0.03. 

5. The underlying power-law shows distinct indices for the low energy and high energy
data, -1.90 (with a typical error of 0.03) and -2.2 (with a typical error of 0.1), 
respectively. We find that this is compatible with a cutoff power-law model for the
non-thermal emission with the power-law index of -1.78$\pm$0.03 and cut-off
energy of $\sim170\pm44$\,keV.

6. Attributing the blackbody emission to photospheric emission, we find that the $\cal{R}$ parameter
of the lower-temperature blackbody is 65$\pm$8 times higher than that of the 
higher-temperature blackbody. The radius of both the photospheres increase monotonically but very slowly with time.

7. If various components of the spectrum are produced due to a spine-sheath jet structure, 
we find that the parameters and their evolution (e.g., ratio of the bulk \emph{Lorentz} factor of the 
two components, temperature and flux evolution etc.) are consistent with the cocoon model.

\section*{Acknowledgments} This research has made use of data obtained through the
HEASARC Online Service, provided by the NASA/GSFC, in support of NASA High Energy
Astrophysics Programs. This work made use of data supplied by the UK Swift Science 
Data Centre at the University of Leicester. We thank the referee specially for 
the valuable comments in re-organizing the presentation of the paper.

\bibliographystyle{apj}
\bibliography{thesis_bib}

\end{document}